\documentclass[aps,prl,twocolumn,showpacs,superscriptaddress,floats]{revtex4-1}
\usepackage{graphicx,bm,amsmath,amssymb,natbib,url,epsfig}
\usepackage{dcolumn}
\usepackage{hyperref}

\usepackage{color}
\definecolor{darkgreen}{rgb}{0.1,.6,.1}
\definecolor{greenblue}{rgb}{0.0,.1,.4}

\usepackage[normalem]{ulem}

\begin{document}

\title{Chimera-Like Coexistence of Synchronized Oscillation and Death in an 
Ecological Network} 

\author{Partha Sharathi Dutta}
\email{parthasharathi@iitrpr.ac.in}
\affiliation{Department of Mathematics, Indian Institute of Technology Ropar,
Rupnagar 140 001, Punjab, India.}
\author{Tanmoy Banerjee}
\thanks{Corresponding author}
\email{tbanerjee@phys.buruniv.ac.in}
\affiliation{Department of Physics, University of Burdwan, Burdwan 713
  104, West Bengal, India.}

\received{:to be included by reviewer}
\date{\today}

\begin{abstract}
We report a novel spatiotemporal state, namely the chimera-like
incongruous coexistence of {\it synchronized oscillation} and {\it
  stable steady state} (CSOD) in a realistic ecological network of
nonlocally coupled oscillators. Unlike the {\it chimera} and
{\it chimera death} state, in the CSOD state identical oscillators are
self-organized into two coexisting spatially separated domains: In one
domain neighboring oscillators show synchronized oscillation and in
another domain the neighboring oscillators randomly populate either a
synchronized oscillating state or a stable steady state (we call it a
death state). We show that the interplay of nonlocality and coupling
strength results in two routes to the CSOD state: One is from a
coexisting mixed state of amplitude chimera and death, and another one is from a globally synchronized state. We
further explore the importance of this study in ecology that gives
a new insight into the relationship between spatial
synchrony and global extinction of species.
\end{abstract}

\maketitle 

Understanding of collective dynamical behaviors in networks of coupled oscillators has been an active area of
extensive research in the field of physics, chemistry, biology,
engineering and social sciences. Coupled oscillators show a
plethora of cooperative phenomena, such as synchronization
\cite{sync}, amplitude death \cite{adrev}, oscillation death
\cite{kosprep}, chimera \cite{chireview}, chimera death \cite{scholl},
etc. Two intriguing spatiotemporal dynamical states, namely the {\it chimera} and the recently observed {\it chimera death} have been in the center of recent research on coupled oscillators for their rich complex behaviors.

The chimera state is a fascinating spatiotemporal state where synchronous and asynchronous oscillations coexist in a
network of coupled identical oscillators. After its
discovery by \citet{kuro} and mathematical proof in
Ref.~\cite{st1,*st2}, the chimera state attracts immediate attention
due to its possible connection with unihemispheric sleep in certain species \cite{chireview}, the multiple time-scales of sleep dynamics \cite{neurochi}, etc. Unlike phase chimera, where chimera occurs in the phase part, recently it is found that in the strong coupling limit amplitude effects come into
play that results in amplitude mediated chimera
\cite{sethia1,*sethia2} and amplitude chimera \cite{scholl}, 
\cite{scholl2,*scholl3}. The existence of chimera has also been established
in many experiments, e.g., in optical system \cite{raj}, chemical
oscillators \cite{show}, mechanical system \cite{mech}, and electronic
system \cite{fortuna}. Further, chimera state has been observed in
various fields; examples include \cite{chiex1,*chiex2,*chiex3}: FitzHugh-Nagumo oscillator, the
SNIPER model of excitability of type-I, autonomous Boolean networks, etc
(for an elaborate review please see \cite{chireview}). Recently,
chimera state in population dynamics is observed using Lattice Limit
Cycle (LLC) model \cite{scholleco}.

On the other hand, the {\it chimera death} (CD) state is discovered very recently by \citet{scholl} in a network of Stuart-Landau oscillators under nonlocal coupling. The CD state connects the chimera state to the oscillation death (OD) state \cite{kosprl}. In the OD state oscillators populate different branches of a stable inhomogeneous steady state (IHSS) \cite{kosprep,*tanpre1,*tanpre2,*tanpre3}. According to Ref.\cite{scholl}, in the {\it chimera death} state the population of oscillators in a network splits into coexisting domains of spatially coherent OD (where neighboring nodes attain essentially the same branch of the IHSS) and spatially incoherent OD (where the neighboring nodes jump among the different branches of IHSS in a completely random manner). Later, CD is also found in a network of mean-field diffusively
coupled oscillators \cite{tanCD}.

In summary, the {\it chimera} state is a spatial coexistence of
coherent and incoherent {\bf oscillations}, and the {\it chimera death} state is
a spatial coexistence of coherent and incoherent branches of {\bf oscillation death} state. Thus, the next natural question arises: {\it Is it possible to have an emergent state in a network of oscillators that shows a chimera-like coexistence of {\bf coherent oscillation} and {\bf stable steady state}}?

In this Letter, for the first time, we indeed find the answer in
affirmative. Here, we address this open question and show that in a
realistic ecological network consists of Rosenzweig--MacArthur
oscillators \cite{RoMa63} under nonlocal coupling topology, the
interplay of non locality and coupling strength gives rise to a novel
spatiotemporal state. In this state, the population of oscillators split into
two coexisting distinct spatially separated domains: In one domain oscillators are
oscillating in synchrony (i.e., coherently), and in another domain
neighboring oscillators depict spatially synchronized oscillation and
stable steady state in a random manner (i.e., incoherently). Hereafter, we call this hitherto unobserved state a {\it
    chimera-like synchronized oscillation and death} (CSOD)
  state (the stable steady state is denoted as a {\it death}
 state). Depending upon the coupling range and coupling strength, we
identify two types of transitions to the CSOD state: With increasing coupling range (for a moderate coupling strength) the CSOD state arises from a coexisting mixed state of amplitude chimera and death state; on the other hand, for an increasing coupling strength (with a moderate coupling range) the CSOD state comes from a globally synchronized oscillation state. However, in both the cases, under large coupling range and strength, the CSOD state is transformed into a chimera death
state. {\it Thus, the CSOD state bridges the gap between the amplitude
  chimera and the chimera death state}. We further discuss the
ecological importance of this emergent behavior that gives us a new
insight into the relationship between spatial synchrony and
global extinction of species, which are thought of as closely
connected phenomena in ecology
\cite{HeKaRa97,*EaRoGr98,*EaLeRo00,*LiKoBj04,*gr}: Spatial
  synchrony may lead to global extinction of species. Here
we show that our results differ from this general consensus, and local extinction of a species does not necessarily lead to a global extinction of that species. 

\begin{figure}
\centering
\includegraphics[width=0.46\textwidth]{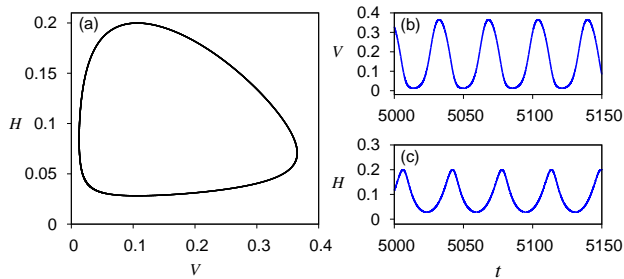}
\caption{\label{fig1} (Color online) (a) Limit cycle attractor, and
  (b), (c) time-series of the uncoupled Rosenzweig--MacArthur model
  given by Eqs.~(\ref{eq1}) for $r= 0.5$, $K=0.5$, $\alpha=1$, $B=
  0.16$, $\beta=0.5$ and $m=0.2$.}
\end{figure}

Here we consider the following network of $N$ identical nonlocally coupled
Rosenzweig--MacArthur (RM) oscillators:
\begin{subequations}\label{eq1}
\begin{align}
\frac{dV_i}{dt}&=rV_i\left(1-\frac{V_i}{K}\right)-\frac{\alpha V_i}{V_i+B}H_i,\\
\frac{dH_i}{dt}&=H_i\left(\frac{\alpha\beta V_i}{V_i+B}-m\right)+\frac{\sigma}{2P}\sum_{k=i-P}^{i+P}\left(H_k-H_i\right),
\end{align}
\end{subequations}
with $V$ and $H$, respectively, representing vegetation and
  herbivore density, interacting in $i~(=1, 2,\cdots, N)$ discrete patches (or nodes) (``$i$'' is taken as modulo
$N$). The local dynamics in each node are governed by the following system parameters: $r$ is the
  intrinsic growth rate, $K$ is the carrying capacity, $\alpha$ is the maximum
  predation rate of the herbivore, $B$ is the half
  saturation constant, $\beta$ represents the herbivore
  efficiency and $m$ is the mortality rate of the
  herbivore.  Interaction between nodes
  is governed by two coupling parameters: $\sigma$ is the coupling strength and $P$ controls
the coupling range, where $1\le P \le \frac{N}{2}$. Two limiting
values of $P$, i.e., $P=1$ and $P=N/2$ represent local and global
coupling, respectively. This nonlocal coupling topology has been used
in Ref.~\cite{scholl} and Ref.~\cite{scholl2,*scholl3} to observe
chimera states in periodic and chaotic oscillators. The Rosenzweig--MacArthur model perhaps
is the simplest model that can actually be applied in real
ecosystems. As a result, this model becomes a standard spatially
  structured prey--predator model in theoretical ecology \cite{HoHa08,*GoHa09,*GoHa11,bandutta}.

An individual RM oscillator [see Eqs.~(\ref{eq1}) for
$\sigma=0$ and a fixed $i$] has the following steady
states: (i) $(V^*,H^*)=(0,0)$, the eigenvalues are $\lambda=r$, $-m$
and the equilibrium point is a saddle point, (ii) $(V^*,H^*)=(K,0)$,
the eigenvalues are $\lambda=-r$, $-m +\alpha\beta \frac{K}{K+B}$ and
the equilibrium point is either a stable node or a saddle node,
depending upon the values of the parameters, and finally (iii)
$(V^*,H^*)=(\frac{mB}{\alpha\beta-m},
\frac{r}{\alpha}(1-\frac{mB}{K(\alpha\beta-m)})
(\frac{B\alpha\beta}{\alpha\beta-m}))$; this nontrivial equilibrium
point is stable for parameter values satisfying the
inequality $\frac{B}{K} < \frac{(\alpha\beta-m)}{(\alpha\beta + m)}$. Beyond a certain $K$, this equilibrium point becomes unstable and gives rise to a stable limit cycle through Hopf bifurcation. In general, further increase in $K$ gradually increases amplitude of the limit cycle, thus bringing the density of either the prey or the predator or both the populations closer to zero, eventually leading to the extinction of the ecosystem; this is known as ``the paradox of enrichment" \cite{Ro71} proposed by  M. L. Rosenzweig in 1971. A subsequent realistic range \cite{Mu98} of $K$ is $0.15$ to $3$ and
range of $m$ is $0.03$ to $0.41$. In Fig.~\ref{fig1}, a stable
limit cycle is shown for the following parameter values: $r= 0.5$,
$K=0.5$, $\alpha=1$, $B= 0.16$, $\beta=0.5$ and $m=0.2$.
\begin{figure}
\centering
\includegraphics[width=0.47\textwidth]{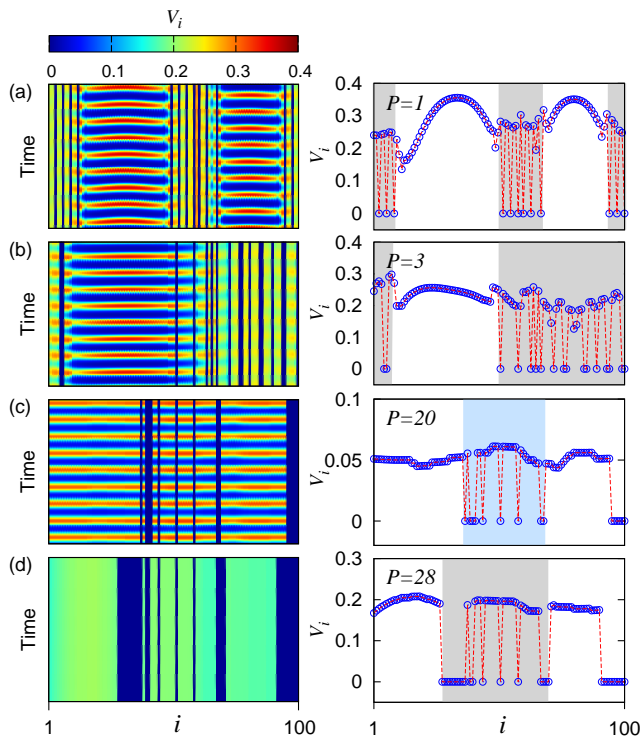}
\caption{\label{fig2} (Color online) Left panel: Spatiotemporal color
  map and right panel: $V_i$ with
  oscillator index ``$i$'' [red (dotted) line is for visual guidance]. Coupling strength $\sigma=1.7$,
  $N=100$. (a) $P=1$ ($\gamma=0.01$) and (b) $P=3$ ($\gamma=0.03$)
  show the mixed state of amplitude chimera and stable zero steady
  state (gray shade in right panels are for visual guidance). (c) $P=20$ ($\gamma=0.2$): {\it Chimera-like synchronized
    oscillation and death} (CSOD) state. Cyan (gray) shaded region in
  the right panel shows the random sequential occurrence of the
  synchronized oscillation and zero steady state of the neighboring
  nodes. (d) $P=28$ ($\gamma=0.28$): The chimera death state. Initial
  time $t=5000$ is discarded before presenting the figures. Other
  parameters are same as used in Fig.\ref{fig1}.}
\end{figure}

To explore various spatiotemporal patterns in the network, we take
$N=100$ and integrate Eqs.~\eqref{eq1} numerically \footnote{We use
  the fourth-order Runge-Kutta algorithm with step size $0.01$}. While
presenting the simulation results, a large number of initial
integration time ($t=5000$) is discarded in order to ensure the steady
state behavior. At first, we consider a moderate coupling strength,
$\sigma=1.7$, and increase the coupling range, $\gamma=P/N$, from a
lower value. For lower coupling range ($\gamma\le0.05$) we observe a
mixed state comprises of amplitude chimera and stable zero steady
state (i.e., death state). This is shown in Figs.~\ref{fig2}(a) and
\ref{fig2}(b) for $\gamma=0.01$ and $\gamma=0.03$, respectively. The
left panel in Fig.~\ref{fig2} shows the space-time color map of $V_i$
and the right panel shows $V_i$ in the steady state with oscillator
index (``$i$"). The gray shaded regions in Figs.~\ref{fig2}(a)
and \ref{fig2}(b) (right panel) show this mixed state of amplitude
chimera and death: {\it the amplitude chimera interrupted by death
  state is a new observation in the context of coupled
  oscillators}. Further increase in coupling range $\gamma$ results in
the CSOD state where the population of oscillators splits into two
distinct coexisting domains: In one domain the neighboring nodes oscillate in
synchrony while in another domain the neighboring nodes randomly
populate either the synchronized oscillating state or the stable
steady (death) state. This new spatiotemporal state is shown in
Fig.~\ref{fig2}(c) for $\gamma=0.2$ (i.e., $P=20$). Here we see that
in the shaded region [right panel of Fig.~\ref{fig2}(c)] the
neighboring oscillators populate either synchronized oscillation
state or stable zero steady state in a random sequence.
However, in the non shaded region only synchronized oscillation exists
except in the rightmost nodes where few oscillators attain the death
state. To the best of our knowledge, this chimera-like spatiotemporal
state is new in literature because here stable steady state coexists
with synchronized oscillations, which is unlike the chimera state or
chimera death state. We also verify that this CSOD state is preserved for a larger number of nodes, $N$   \footnote{See Supplemental Material at [URL will be inserted by publisher] for the CSOD state for $N=300$}. If we further increase the coupling range, we
find the chimera death state [Fig.~\ref{fig2}(d) for
  $\gamma=0.28$]. We notice that instead of populating {\it two}
branches of OD \cite{scholl}, \cite{tanCD}, denoted as lower and upper
branch, here in the CD state, the OD state has more than two branches [later it is clearly shown in Fig.~\ref{fig5}(b)]. The multiple-branches (more than two) of OD was reported earlier in \cite{ref82} for sixteen {\it locally} coupled
genetic relaxation oscillators, but in a network of large number of 
oscillators with {\it nonlocal} coupling it is an important
observation (to be discussed later on).

\begin{figure}
\centering
\includegraphics[width=0.45\textwidth]{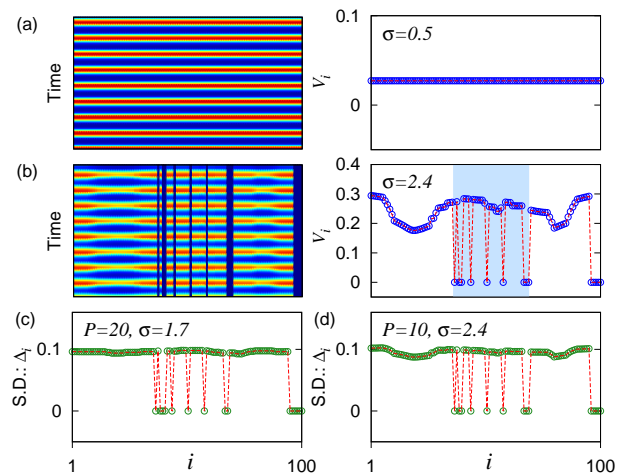}
\caption{\label{fig3} (Color online) Left panel: Spatiotemporal color
  map (color bar same as Fig.~\ref{fig2}), and right panel: $V_i$ with oscillator index ``$i$''. Coupling range
  $\gamma=0.1$ (i.e., $P=10$), $N=100$. (a) $\sigma=0.5$: Global
  in-phase synchronized oscillation. (b) $\sigma=2.4$ : The CSOD
  state. Cyan (gray) shaded region is for visual guidance of the incoherent
  region. (c) Standard deviation (S.D.) ($\Delta_i$) with index
  ``$i$'' for the CSOD state of Fig.~\ref{fig2}(c). (d) $\Delta_i$
  with index ``$i$'' for the CSOD state of Fig.~\ref{fig3}(b). Other
  parameters are same as used in Fig.\ref{fig1}.}
\end{figure}

We also identify one more significant route to the CSOD state
with increasing coupling strength ($\sigma$) and a fixed $\gamma$,
namely the transition from a global in-phase synchronized oscillating state to the
CSOD state. This transition is shown in Figs.~\ref{fig3}(a) and
\ref{fig3}(b) for $\sigma=0.5$ and $\sigma=2.4$,
respectively (for $\gamma=0.1$, i.e., $P=10$). Here also, the CSOD
state is transformed into a multi-branch chimera death state for higher
coupling strength (not shown).

Next, we quantify the spatiotemporal behavior where synchronized
oscillation and stable zero steady (death) state coexist. In order to
distinguish between the oscillation and death, we compute the standard
deviation (S.D.) of each node given by:
\begin{equation}
\displaystyle \Delta_i=\sqrt{\left(<V_{i}^{2}>-<V_i>^2\right)}.
\end{equation}
The ``$<\;>$'' sign denotes the time average, which is carried out over a
long time period ($t=3000$ in the steady state). For a stable steady
state (i.e., a death state) S.D. ($\Delta_i$) must be zero
and in the oscillatory condition it will show a finite non-zero
value. Figures \ref{fig3}(c) and \ref{fig3}(d) show $\Delta_i$ values of
the CSOD states shown in Fig.~\ref{fig2}(c) and Fig.~\ref{fig3}(b),
respectively. For the CSOD state, in the incoherent region, we see that
the $\Delta_i$ changes from a finite non-zero value to zero in a
random manner; in the populations where nodes are oscillating in synchrony its value is non-zero and shows a continuous spatial variation.

\begin{figure}
\centering
\includegraphics[width=0.38\textwidth]{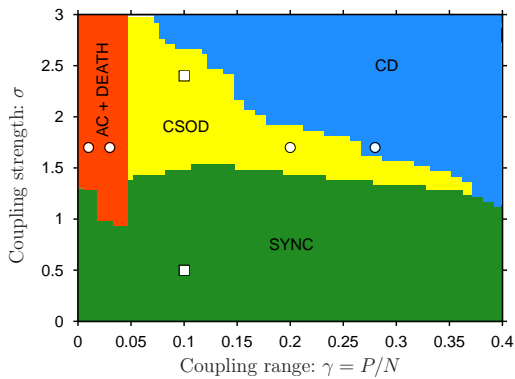}
\caption{\label{fig4} (Color online) Phase diagram in the
  $\gamma-\sigma$ space. {\bf CSOD}: chimera-like synchronized
  oscillation and stable zero steady state (death); {\bf SYNC}: Global
  in-phase synchronized oscillation; {\bf CD}: Chimera death; {\bf
    AC+Death}: Coexistence of amplitude chimera and stable zero steady
  state. The symbols $\circ$ indicate the coupling parameter values
  used for generating Figs.~\ref{fig2} (a)-\ref{fig2}(d). $\Box$
  represents the same for generating Figs.~\ref{fig3}(a) and
  \ref{fig3}(b). Other parameters are same as used in Fig.\ref{fig1}.}
\end{figure}

In order to reveal the complete spatiotemporal scenario of the
considered network, we rigorously compute the phase
diagram in the $\gamma-\sigma$ space (the unsynchronized zone with very small $\sigma$ value is not shown). From the phase diagram
[Fig.~\ref{fig4}] it is clear that the region of occurrence of
the CSOD state is broad enough. It is seen that beyond $\gamma \approx 0.37$ (i.e., $P\approx 37$) no CSOD occurs; here an increase in $\sigma$ transforms the synchronized oscillation state (SYNC) directly to the chimera death. The symbols $\circ$ in the phase diagram indicate the coupling parameter values used for
  generating Figs.~\ref{fig2}(a)-\ref{fig2}(d), whereas $\Box$ represents the
  same for generating Figs.~\ref{fig3}(a) and \ref{fig3}(b).
In this context, it should be noted that in the phase diagram the boundaries
among different phases are not very sharp, they tend to change with
initial conditions. However, we observe that the overall qualitative
structure of the phase diagram is preserved for all the initial
conditions or number of nodes.

\begin{figure}
\centering
\includegraphics[width=0.35\textwidth]{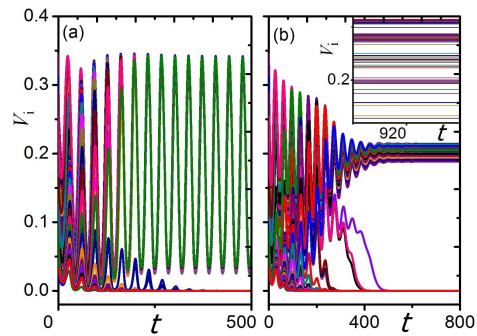}
\caption{\label{fig5} (Color online) Time series of $V_i$'s for (a)
  $\gamma=0.2$ and $\sigma=1.7$: Synchronized limit-cycle oscillation
  (upper branch) and stable zero steady state (i.e., $V=0$)
  coexists. (b) $\gamma=0.28$ and $\sigma=1.7$: Temporally stable multi-branch OD; Inset shows the multiple branches around $V_i=0.2$. Other parameters are same as used in Fig.\ref{fig1}.}
\end{figure}

Next, we provide a qualitative explanation of the
genesis of the CSOD state. We find that it has a strong connection
with the {\it inhomogeneous limit-cycle} (IHLC) state in a network
\cite{tyson,*kosihlc1}, \cite{kosihlc}; IHLC is defined as a
state where some nodes are in a stable steady state (or quasisteady
state with a negligible amplitude, as in \cite{kosihlc}), while the
rest undergo oscillations. To visualize the scenario, we plot the time
series of all the $V_i$'s [Fig.~\ref{fig5}(a)] for an exemplary
value of $\gamma=0.2$ and $\sigma=1.7$. From Fig.~\ref{fig5}(a) one
observes that a population of oscillators occupy the trivial zero steady state (i.e., $V=0$
state), while the rest of the oscillators are in the in-phase
synchronized oscillating state. Thus, depending upon judiciously 
chosen {\it spatial} initial conditions, individual
nodes may populate either the upper oscillating branch or the
lower steady state (i.e., $V=0$) branch in a random
sequence, which results in the CSOD state [as shown in Fig.~\ref{fig2}(c)]. For higher coupling range and
strength, e.g., $\gamma=0.28$ and $\sigma=1.7$, oscillators populate the multi-branch OD state [Fig.~\ref{fig5}(b), see also the inset]; here a set of proper spatial initial conditions should result in chimera death in the
network [as shown in Fig.~\ref{fig2}(d)]. Thus, we may conjecture that, the CSOD state may occur in systems where this
type of [Fig.~\ref{fig5}(a)] IHLC state exists.

Finally, we discuss the importance of the results in ecology. In spatial ecology nonlocal coupling arises under the assumption that all spatially separated patches (or nodes) are connected only to certain number of
  neighboring nodes in a fragmented landscape, which is a more natural coupling scheme than the global coupling. In ecology, it is generally believed that spatial synchrony and global extinction are two strongly correlated phenomena (see for example,
Refs.\cite{HeKaRa97,*EaRoGr98,*EaLeRo00,*LiKoBj04,*gr}). In contrast
to this general belief, in the present study we show that, in nonlocal
dispersive coupling, although spatial synchronization gives rise to local extinction of a species in one or more patches (or nodes) [i.e., the {\it death} state] but it defies the {\it global extinction} of the
species (i.e., not all the oscillators go to the {\it death} state). Moreover, a general consensus in ecology is that, spatial
synchrony and dispersal induced stability (or temporal stability) are
two conflicting outcomes of dispersion among the population of
patches. In the existing studies it is shown that dispersion among
identical patches results in spatial synchrony; on the other hand, the
combination of spatial heterogeneity and dispersion is necessary for
dispersal-induced stability \citep{BrHo04,*Ab11}. Here our results show that depending on coupling range and strength, spatial synchronization among {\it identical} patches (or nodes) leads to temporally stable multi-branch (more than two) IHSS and a cluster of them has non-zero steady states. Thus, to
achieve temporal stability the patches need not to be heterogeneous
but nonlocal coupling is sufficient. Further, the occurrence of the amplitude chimera interrupted by death is a new finding and its proper interpretation in
ecology deserves further attention.

In conclusion, in this Letter we have reported a novel spatiotemporal state, the CSOD state, in a realistic ecological network with nonlocal coupling topology. In this state a subset of oscillators populate spatially synchronized oscillation and stable steady state in a random manner, and
the rest of the oscillators oscillate in synchrony. This
spatiotemporal state is unlike the chimera state (where
coherent and incoherent {\it oscillations} coexist) and the chimera
death state (where neighboring oscillators populate two branches of
{\it OD} in coherent and incoherent manner). We have shown two coupling dependent  transition routes to this
CSOD state. We further qualitatively established the connection
of this emergent state with the inhomogeneous limit-cycle state present in the network. We have discussed the ecological importance of the results, which reveals that spatial synchrony does not necessarily lead to global extinction of a species, which is in contrast to the general consensus. Apart from ecology, we believe that, the present study will improve our understanding of other physical networks, e.g., power grid and communication networks, where it is desirable that a failure of certain nodes does not lead to a complete blackout or a global system failure \cite{grid1,*grid2}. 

T.B. acknowledges financial support from SERB, Department of Science
and Technology (DST), India (Grant No. SB/FTP/PS-005/2013).

\providecommand{\noopsort}[1]{}\providecommand{\singleletter}[1]{#1}%
\end{document}